\pdfoutput=1

\documentclass[12pt,a4paper]{article}

\usepackage{ifthen} 
\newboolean{pdflatex}
\setboolean{pdflatex}{true} 

\newboolean{articletitles}
\setboolean{articletitles}{true} 

\newboolean{uprightparticles}
\setboolean{uprightparticles}{false} 

\newboolean{inbibliography}
\setboolean{inbibliography}{false} 

\usepackage{microtype}
\usepackage{lineno}  
\usepackage{xspace} 

\usepackage{graphicx}  
\usepackage{color}
\usepackage{colortbl}
\graphicspath{{./figs/}} 

\usepackage{amsmath} 
\usepackage{amssymb}
\usepackage{amsfonts}
\usepackage{upgreek} 

\newcommand*\patchAmsMathEnvironmentForLineno[1]{%
\expandafter\let\csname old#1\expandafter\endcsname\csname #1\endcsname
\expandafter\let\csname oldend#1\expandafter\endcsname\csname
end#1\endcsname
 \renewenvironment{#1}%
   {\linenomath\csname old#1\endcsname}%
   {\csname oldend#1\endcsname\endlinenomath}%
}
\newcommand*\patchBothAmsMathEnvironmentsForLineno[1]{%
  \patchAmsMathEnvironmentForLineno{#1}%
  \patchAmsMathEnvironmentForLineno{#1*}%
}
\AtBeginDocument{%
\patchBothAmsMathEnvironmentsForLineno{equation}%
\patchBothAmsMathEnvironmentsForLineno{align}%
\patchBothAmsMathEnvironmentsForLineno{flalign}%
\patchBothAmsMathEnvironmentsForLineno{alignat}%
\patchBothAmsMathEnvironmentsForLineno{gather}%
\patchBothAmsMathEnvironmentsForLineno{multline}%
}

\usepackage{hyperref}    
\usepackage[all]{hypcap} 

\def\ux85 {\mbox{UX85}\xspace}

\ifthenelse{\boolean{uprightparticles}}%
{

 \def\Ppi         {\ensuremath{\uppi}\xspace}

 \def\Ppsi        {\ensuremath{\uppsi}\xspace}

 \def\PDelta      {\ensuremath{\Delta}\xspace}                 
 \def\PXi      {\ensuremath{\Xi}\xspace}                 
 \def\PLambda      {\ensuremath{\Lambda}\xspace}                 
 \def\PSigma      {\ensuremath{\Sigma}\xspace}                 
 \def\POmega      {\ensuremath{\Omega}\xspace}                 
 \def\PUpsilon      {\ensuremath{\Upsilon}\xspace}                 
 
 \def\PB      {\ensuremath{\mathrm{B}}\xspace}                 
                  
 \def\PD      {\ensuremath{\mathrm{D}}\xspace}

 \def\PJ      {\ensuremath{\mathrm{J}}\xspace}                 
 \def\PK      {\ensuremath{\mathrm{K}}\xspace}

 \def\Pb      {\ensuremath{\mathrm{b}}\xspace}

 \def\Pi      {\ensuremath{\mathrm{i}}\xspace}

 \def\Ps      {\ensuremath{\mathrm{s}}\xspace}

}
{

 \def\Ppi         {\ensuremath{\pi}\xspace}

 \def\Ppsi        {\ensuremath{\psi}\xspace}                 
                  
 \mathchardef\PDelta="7101
 \mathchardef\PXi="7104
 \mathchardef\PLambda="7103
 \mathchardef\PSigma="7106
 \mathchardef\POmega="710A
 \mathchardef\PUpsilon="7107
                  
 \def\PB      {\ensuremath{B}\xspace}                 
                  
 \def\PD      {\ensuremath{D}\xspace}

 \def\PJ      {\ensuremath{J}\xspace}                 
 \def\PK      {\ensuremath{K}\xspace}

 \def\Pb      {\ensuremath{b}\xspace}

 \def\Pi      {\ensuremath{i}\xspace}

 \def\Ps      {\ensuremath{s}\xspace}

}

\def\squark    {\ensuremath{\Ps}\xspace}

\def\bquark    {\ensuremath{\Pb}\xspace}

\def\pion  {\ensuremath{\Ppi}\xspace}

\def\pip   {\ensuremath{\pion^+}\xspace}
\def\pim   {\ensuremath{\pion^-}\xspace}
\def\pipi  {\ensuremath{\pion^+\pion^-}\xspace}

\def\kaon  {\ensuremath{\PK}\xspace}
  \def\Kbar  {\kern 0.2em\overline{\kern -0.2em \PK}{}\xspace}

\def\Kz    {\ensuremath{\kaon^0}\xspace}
\def\Kzb   {\ensuremath{\Kbar^0}\xspace}
\def\KzKzb {\ensuremath{\Kz \kern -0.16em \Kzb}\xspace}
\def\Kp    {\ensuremath{\kaon^+}\xspace}
\def\Km    {\ensuremath{\kaon^-}\xspace}

\def\KpKm  {\ensuremath{\Kp \kern -0.16em \Km}\xspace}

\def\Kstarz  {\ensuremath{\kaon^{*0}}\xspace}
\def\Kstarzb {\ensuremath{\Kbar^{*0}}\xspace}

  \def\Dbar    {\kern 0.2em\overline{\kern -0.2em \PD}{}\xspace}
\def\D       {\ensuremath{\PD}\xspace}

\def\Dz      {\ensuremath{\D^0}\xspace}
\def\Dzb     {\ensuremath{\Dbar^0}\xspace}
\def\DzDzb   {\ensuremath{\Dz {\kern -0.16em \Dzb}}\xspace}
\def\Dp      {\ensuremath{\D^+}\xspace}
\def\Dm      {\ensuremath{\D^-}\xspace}

\def\DpDm    {\ensuremath{\Dp {\kern -0.16em \Dm}}\xspace}

\def\B       {\ensuremath{\PB}\xspace}
\def\Bbar    {\ensuremath{\kern 0.18em\overline{\kern -0.18em \PB}{}}\xspace}

\def\Bzb     {\ensuremath{\Bbar^0}\xspace}

\def\Bd      {\ensuremath{\B^0}\xspace}
\def\Bs      {\ensuremath{\B^0_\squark}\xspace}
\def\Bsb     {\ensuremath{\Bbar^0_\squark}\xspace}
\def\Bdb     {\ensuremath{\Bbar^0}\xspace}

\def\jpsi     {\ensuremath{{\PJ\mskip -3mu/\mskip -2mu\Ppsi\mskip 2mu}}\xspace}

  \def\Y#1S{\ensuremath{\PUpsilon{(#1S)}}\xspace}

\def\L {\ensuremath{\PLambda}\xspace}

\def\Lbar {\ensuremath{\kern 0.1em\overline{\kern -0.1em\PLambda}}\xspace}

\def\Lb      {\ensuremath{\L^0_\bquark}\xspace}

\newcommand{\decay}[2]{\ensuremath{#1\!\to #2}\xspace}         

\def\to                 {\ensuremath{\rightarrow}\xspace}

\def\CP                {\ensuremath{C\!P}\xspace}

\def\AT#1     {\ensuremath{A_{\mathrm{T}}^{#1}}\xspace}           

\def\C#1      {\ensuremath{\mathcal{C}_{#1}}\xspace}                       
\def\Cp#1     {\ensuremath{\mathcal{C}_{#1}^{'}}\xspace}                    
\def\Ceff#1   {\ensuremath{\mathcal{C}_{#1}^{\mathrm{(eff)}}}\xspace}        
\def\Cpeff#1  {\ensuremath{\mathcal{C}_{#1}^{'\mathrm{(eff)}}}\xspace}       
\def\Ope#1    {\ensuremath{\mathcal{O}_{#1}}\xspace}                       
\def\Opep#1   {\ensuremath{\mathcal{O}_{#1}^{'}}\xspace}                    

\newcommand{\tev}{\ifthenelse{\boolean{inbibliography}}{\ensuremath{~T\kern -0.05em eV}\xspace}{\ensuremath{\mathrm{\,Te\kern -0.1em V}}\xspace}}
\newcommand{\gev}{\ensuremath{\mathrm{\,Ge\kern -0.1em V}}\xspace}
\newcommand{\mev}{\ensuremath{\mathrm{\,Me\kern -0.1em V}}\xspace}
\newcommand{\kev}{\ensuremath{\mathrm{\,ke\kern -0.1em V}}\xspace}
\newcommand{\ev}{\ensuremath{\mathrm{\,e\kern -0.1em V}}\xspace}
\newcommand{\gevc}{\ensuremath{{\mathrm{\,Ge\kern -0.1em V\!/}c}}\xspace}
\newcommand{\mevc}{\ensuremath{{\mathrm{\,Me\kern -0.1em V\!/}c}}\xspace}
\newcommand{\gevcc}{\ensuremath{{\mathrm{\,Ge\kern -0.1em V\!/}c^2}}\xspace}
\newcommand{\gevgevcccc}{\ensuremath{{\mathrm{\,Ge\kern -0.1em V^2\!/}c^4}}\xspace}
\newcommand{\mevcc}{\ensuremath{{\mathrm{\,Me\kern -0.1em V\!/}c^2}}\xspace}

\def\gsim{{~\raise.15em\hbox{$>$}\kern-.85em
          \lower.35em\hbox{$\sim$}~}\xspace}
\def\lsim{{~\raise.15em\hbox{$<$}\kern-.85em
          \lower.35em\hbox{$\sim$}~}\xspace}

\def\tell1  {TELL1\xspace}
\def\ukl1   {UKL1\xspace}

\usepackage{cite} 
\usepackage{mciteplus}

\usepackage{longtable} 

\begin{document}

\renewcommand{\thefootnote}{\fnsymbol{footnote}}
\setcounter{footnote}{1}

\begin{titlepage}
\pagenumbering{roman}

\vspace*{-1.5cm}
\vspace*{1.5cm}
\hspace*{-5mm}\begin{tabular*}{16cm}{lc@{\extracolsep{\fill}}r}
 & & DPF2013-137\\
 & & September 23, 2013 \\ 
 & & \\
\end{tabular*}

\vspace*{3.0cm}

{\bf\boldmath\huge
\begin{center}
 Measurements of $b$-hadron lifetimes and effective lifetimes at LHCb
\end{center}
}

\vspace*{2.0cm}

\begin{center}
Bilas Pal\footnote{bkpal@syr.edu}, Syracuse University\\
On behalf of the LHCb collaboration\\

\end{center}

\vspace{\fill}

\begin{abstract}
  \noindent
Precision measurements of $b$-hadron lifetimes are a key goal of the LHCb experiment. In the $\Bs$ sector, the measurement of the effective lifetimes for $\Bs$ mesons decaying to CP-odd, CP-even and flavor specific final states are essential for constraining the $\Bs$ mixing parameters, $\Delta \Gamma_s$, the average width, $\Gamma_s$, and the CP-violating phase, $\phi_s$. Measurements of $b$ baryon lifetimes are also important to test theoretical models. We present the latest results from LHCb on these topics.
\end{abstract}

\vspace*{2.0cm}
\begin{center}
{\large Presented at}
\end{center}
\begin{center}
  DPF 2013\\
The Meeting of the American Physical Society\\
Division of Particles and Fields\\
Santa Cruz, California, August 13--17, 2013\\
\end{center}

\vspace{\fill}

\vspace*{2mm}

\end{titlepage}

\newpage
\setcounter{page}{2}




\renewcommand{\thefootnote}{\arabic{footnote}}
\setcounter{footnote}{0}



\pagestyle{plain} 
\setcounter{page}{1}
\pagenumbering{arabic}  
\section{Introduction}
\noindent
In the free quark model the lifetimes of all $b$-flavored hadrons are equal, because the decay width is determined by the $b$ quark lifetime. This model is too na\"ive, since effects of other quarks in the hadron are not taken into account. Early predictions using the heavy quark expansion (HQE), however, supported the idea that $b$ hadron lifetimes were quite similar, due to the absence of correction terms ${\cal{O}}(1/m_b)$~\cite{Shifman:1986mx,Rosner:1996fy,Neubert:1996we, Cheng:1997xba, Uraltsev:1998bk}. In the case of the ratio of $b$ hadron lifetimes, the corrections of order ${\cal{O}}(1/m_b^2)$ were found to be small, initial estimates of ${\cal{O}}(1/m_b^3)$~\cite{Uraltsev:1996ta,Neubert:1996we,DiPierro:1999tb} effects were also small, thus, differences of only a few percent were expected~\cite{Rosner:1996fy,Neubert:1996we,Cheng:1997xba}.  HQE predicts the ratios of lifetimes of $b$ mesons, which agree the experimental observations well within the experimental and theoretical uncertainties.  The low experimental value of the ratio of lifetimes of the \Lb baryon, $\tau_{\Lb}$, to the \Bzb meson, $\tau_{\Bzb}$, has long been a problem for the theory. This situation has recently been clarified by the LHCb experiment and lends substantial support to  HQE theory.

In these proceedings, we present measurements of $\Gamma_s$ and $\Delta \Gamma_s$ from $\Bs\to\jpsi \phi$ and combined results from $\Bs\to\jpsi \phi$ and $\Bsb\to\jpsi\pip\pim$, the  effective lifetimes of $\Bs$ 
from the decay channels $\Bs\to\KpKm$ and $\Bsb\to\jpsi f_0(980)$ and the lifetime of the $\Lb$ from the $\Lb\to\jpsi p\Km$ decay.

\section{Measurements of $\Gamma_s$ and $\Delta \Gamma_s$}  
\noindent
The decay $\Bs\to\jpsi\phi$ is the golden decay mode for the measurement of the  \CP violation phase, $\phi_s$. The final state is an admixture of the \CP-even and \CP-odd eigenstates. Therefore, an angular analysis is required to disentangle statistically between the two final states with two different \CP eigenvalue. On the other hand, the decay  $\Bs\to\jpsi\pip\pim$ is  shown to be an almost pure \CP-odd eigenstate~\cite{LHCb:2012ae}, and hence no angular analysis is required. Using $1.0~\rm fb^{-1}$ of data collected in $pp$ collisions at $\sqrt{s} = 7$ TeV with the LHCb detector, the \CP violating phase $\phi_s$ as well as $\Gamma_s$ and $\Delta \Gamma_s$ are  extracted by performing an unbinned maximum log-likelihood fit  to the $\Bs$ mass, decay time $t$, angular distributions and initial flavour tag of the selected $\Bs\to\jpsi\phi$ events. The results obtained from the fits are: $\Gamma_s=0.663\pm0.005\pm0.006~\rm ps^{-1}$ and $\Delta \Gamma_s=0.100\pm0.016\pm0.003~\rm ps^{-1}$~\cite{Aaij:2013oba}. The result of performing a combined fit using both $\Bs\to\jpsi\phi$ and $\Bs\to\jpsi\pip\pim$ events gives $\Gamma_s=0.661\pm0.004\pm0.006~\rm ps^{-1}$ and $\Delta \Gamma_s=0.106\pm0.011\pm0.007~\rm ps^{-1}$~\cite{Aaij:2013oba}. These are the most precise measurements to date and are in agreement with the Standard Model (SM) predictions~\cite{Lenz:2006hd, Badin:2007bv, Lenz:2011ti, Charles:2011va}. Since the mean $\Bs$ lifetime defined as $\tau_{\B_s}=1/\Gamma_s$, the measurement of $\Gamma_s$ can be translated into the measurement of $\tau_{\Bs}$ and dividing it with the mean  $\Bd$ lifetime~\cite{PDG}, we obtain the ratio of lifetimes, $\tau_{\Bs}/\tau_{\Bd}=0.996\pm0.006\pm0.010$, which agrees the HQE prediction very well~\cite{Neubert:1996we, Cheng:1997xba, Uraltsev:1998bk}.

\section{Measurement of the $\Bs\to\KpKm$  effective lifetime }
\noindent
A measurement of the effective lifetime of the $\Bs\to\KpKm$ decay is of considerable interest as it is sensitive to new physical phenomena affecting the $\Bs$ mixing phase and entering the decay at loop level. The $\KpKm$ final state is \CP-even and so in the SM the decay is dominated by the light mass eigenstate and the effective lifetime thus is approximately equal to $\Gamma_L^{-1}$, with the assumption that the \CP violation in this decay is small.

Conventional approaches select $\Bs$ meson decay products that are significantly displaced from the primary interaction point. As a consequence, $\Bs$ mesons with low decay times are suppressed, introducing a time-dependent acceptance which  needs to be taken into account. The key principle of this analysis is that the trigger and event selection do not bias
the decay time distribution of the selected $\Bs\to\KpKm$ candidates other than in a trivial way through a minimum decay time requirement.  This has been tested extensively using simulated events at each stage of the selection process to demonstrate that no step introduces a time-dependent acceptance.
Using $1.0~\rm fb^{-1}$ of data recorded in 2011, we measure the effective $\Bs\to\KpKm$ lifetime using a technique  based on neural networks which avoids  acceptance effects.  Only properties independent of the decay time are used to discriminate between signal and background. 
\begin{figure}[htb]
\begin{center}
    \includegraphics[width=0.48\textwidth]{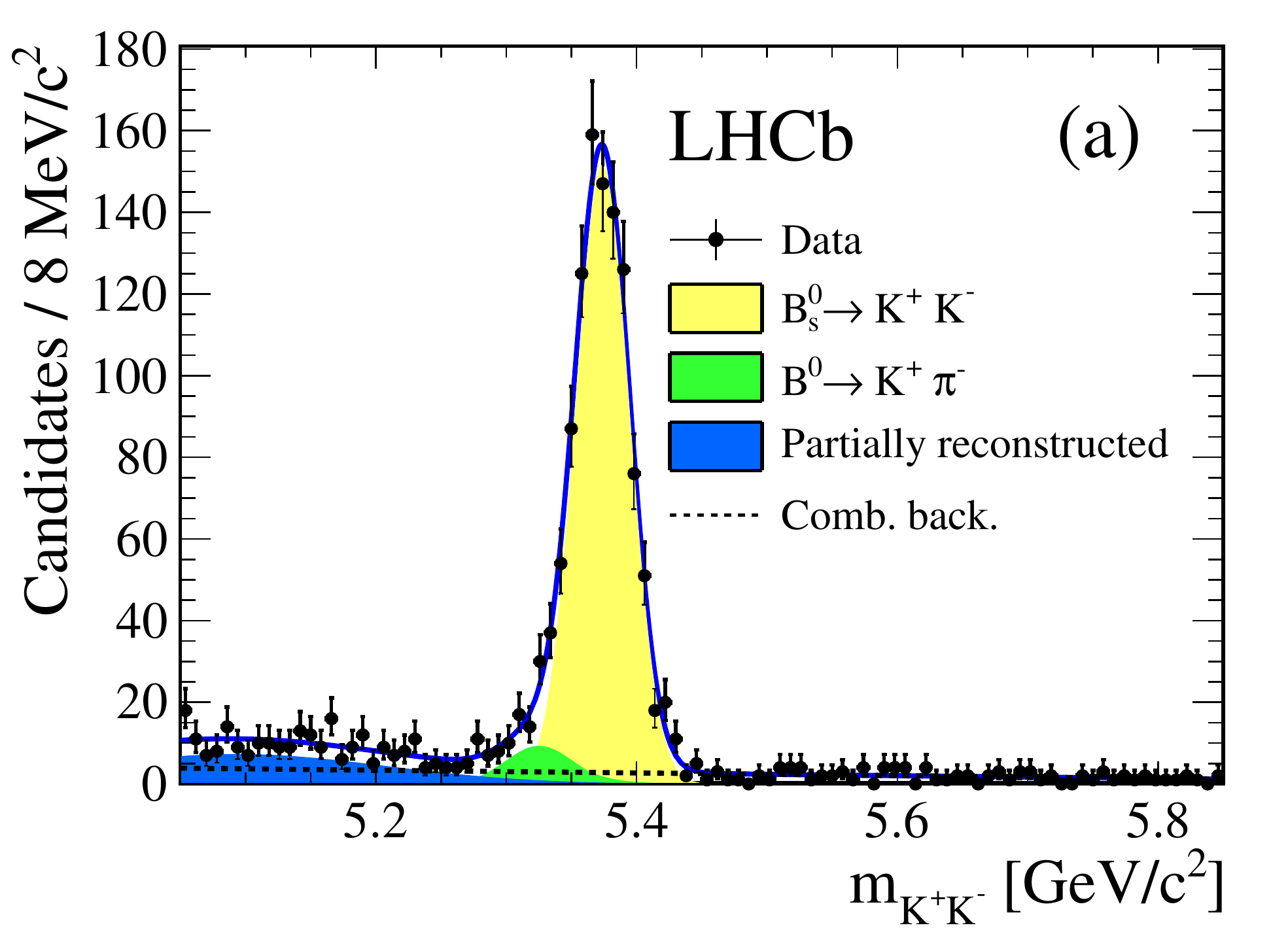}%
    \includegraphics[width=0.48\textwidth]{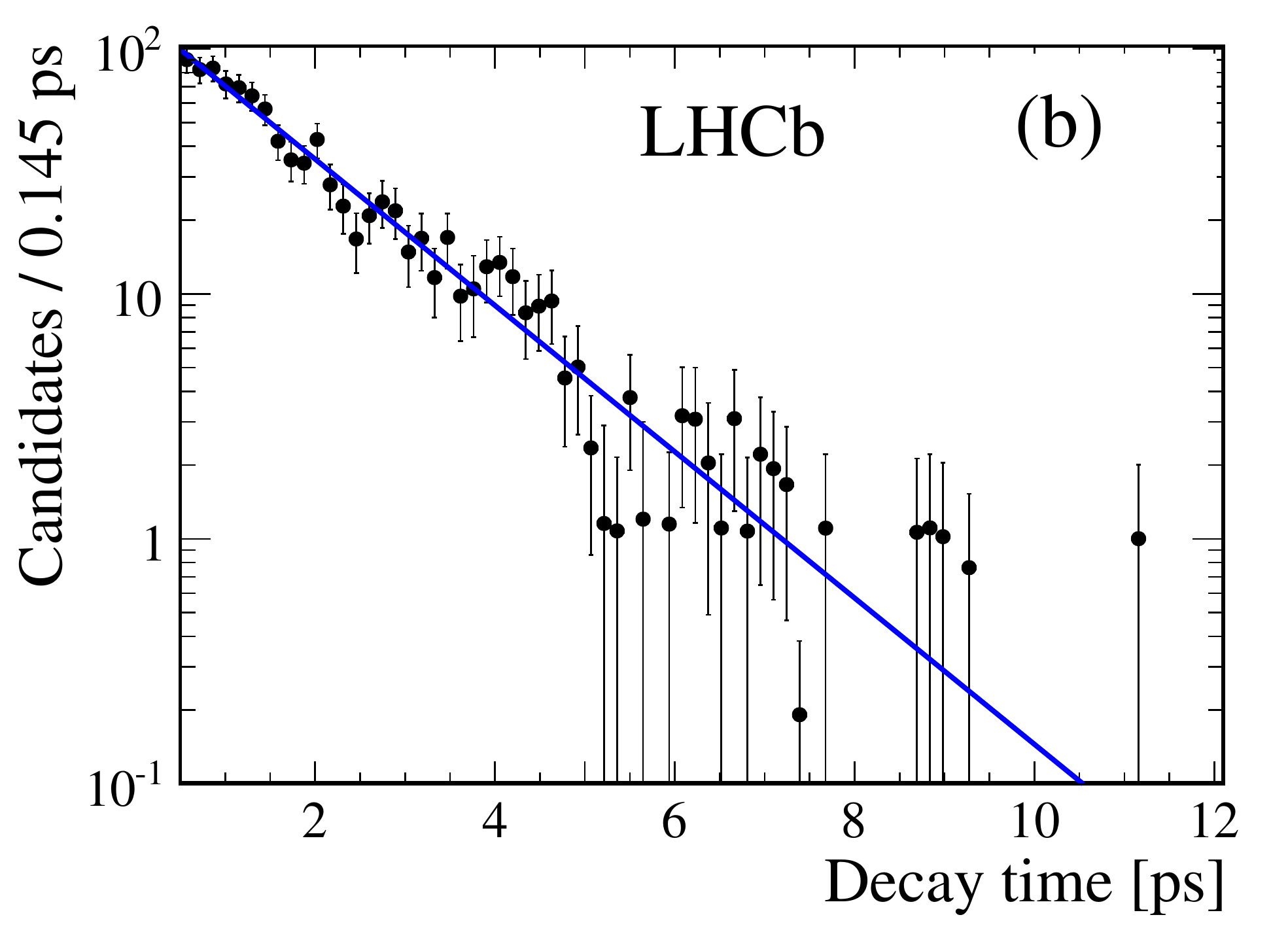}%
\end{center}\label{fig:BsToKK}
\vskip -0.5cm
\caption{\small (a) Invariant mass spectrum for all selected $\Bs\to\KpKm$ candidates. (b) Decay time distribution of $\Bs\to\KpKm$ signal extracted using sWeights and the fitted exponential function.}
\end{figure}

The effective $\Bs\to\KpKm$ lifetime is evaluated using an unbinned
maximum log-likelihood fit to the $\Bs$ decay time distribution. A fit to the invariant mass spectrum is performed to determine the sWeights~\cite{Pivk:2004ty} that are used to isolate the $\Bs\to\KpKm$ decay time distribution from the residual background.  Since there is no acceptance bias to correct for, the lifetime is determined using a fit of the convolution of an exponential and Gaussian function to account for the resolution of the detector. The resolution is determined from the simulation. Both the fit to the invariant mass spectrum and decay time distribution are shown in Fig.~\ref{fig:BsToKK}. The effective  $\Bs\to\jpsi\KpKm$ lifetime is found to be $\tau_{KK}=1.455\pm0.046\pm 0.006~\rm ps$~\cite{Aaij:2012ns} in good agreement with the SM prediction of $1.40\pm0.02~\rm ps$~\cite{Fleischer:2010ib}. The main systematic
contribution is coming from the bias in the lifetime, found after the reconstruction of the tracks in the final state, which is determined from simulated events.

\section{Measurement of the $\Bsb\to\jpsi f_0(980)$  effective lifetime }
\noindent
The $\Bs$ effective lifetime has been determined also in the channel $\Bsb\to\jpsi f_0(980)$. The $\jpsi f_0(980)$ final state is \CP-odd, and in the absence of \CP violation, can be produced only by the decay of the heavy ($\rm{H}$), and not by the light ($\rm{L}$), \Bsb mass eigenstate. As the measured \CP violation in this final state is small~\cite{LHCb:2011ab}, a measurement of the effective lifetime, $\tau_{\jpsi f_0(980)}$, can be translated into a measurement of the decay width, $\Gamma_{\rm H}$. Our procedure involves measuring the lifetime with respect to the well measured \Bdb lifetime, in the decay mode $\Bdb\to\jpsi\Kstarzb(892)$ with $\Kstarzb(892)\to\Km\pip$.  The
advantage of this technique is that in the ratio the decay time acceptance introduced by the trigger, reconstruction and selection requirements and the systematic uncertainties largely cancel.

\begin{figure}[htb]
\begin{center}
   \includegraphics[width=0.48\textwidth]{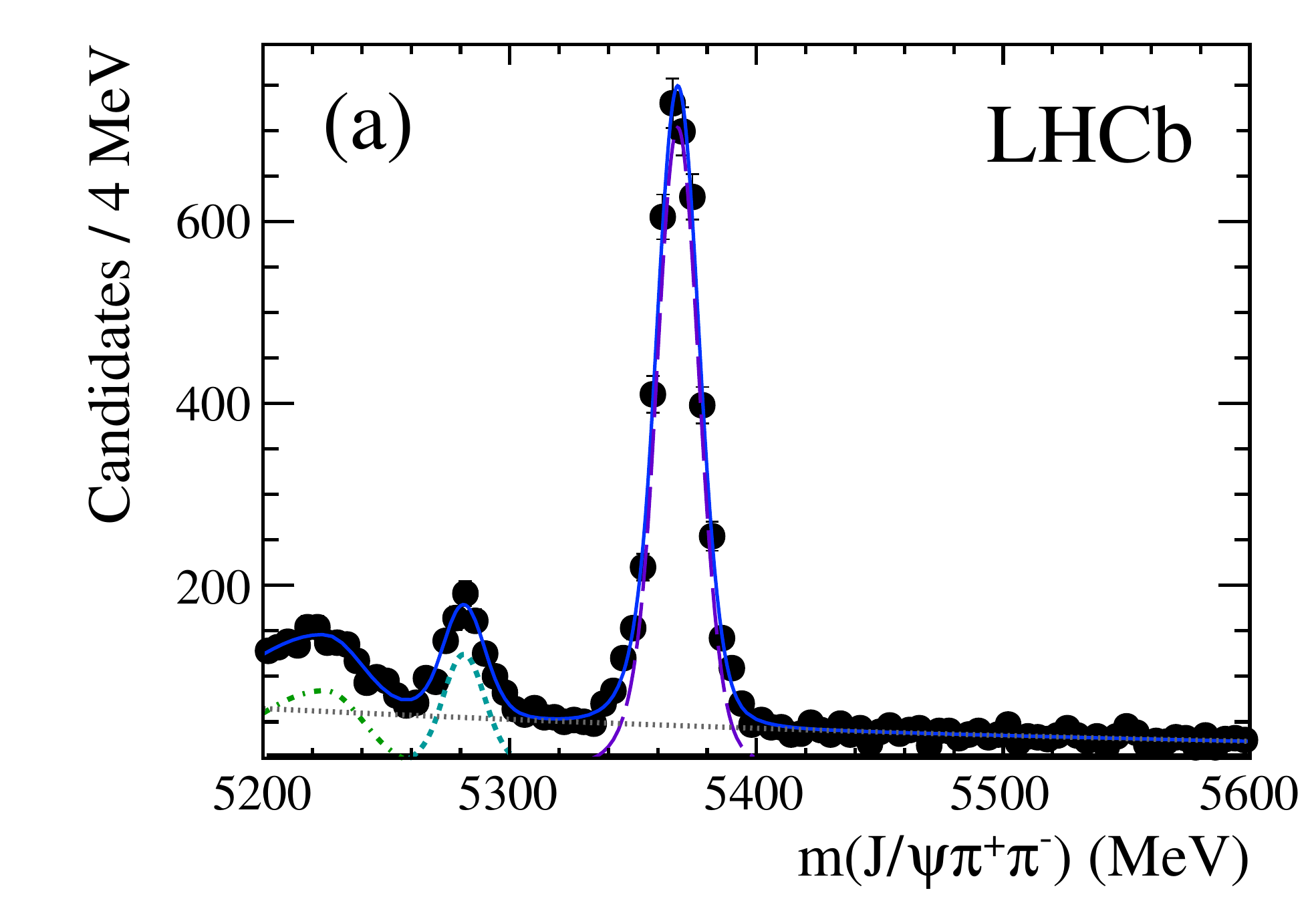}
   \includegraphics[width=0.48\textwidth]{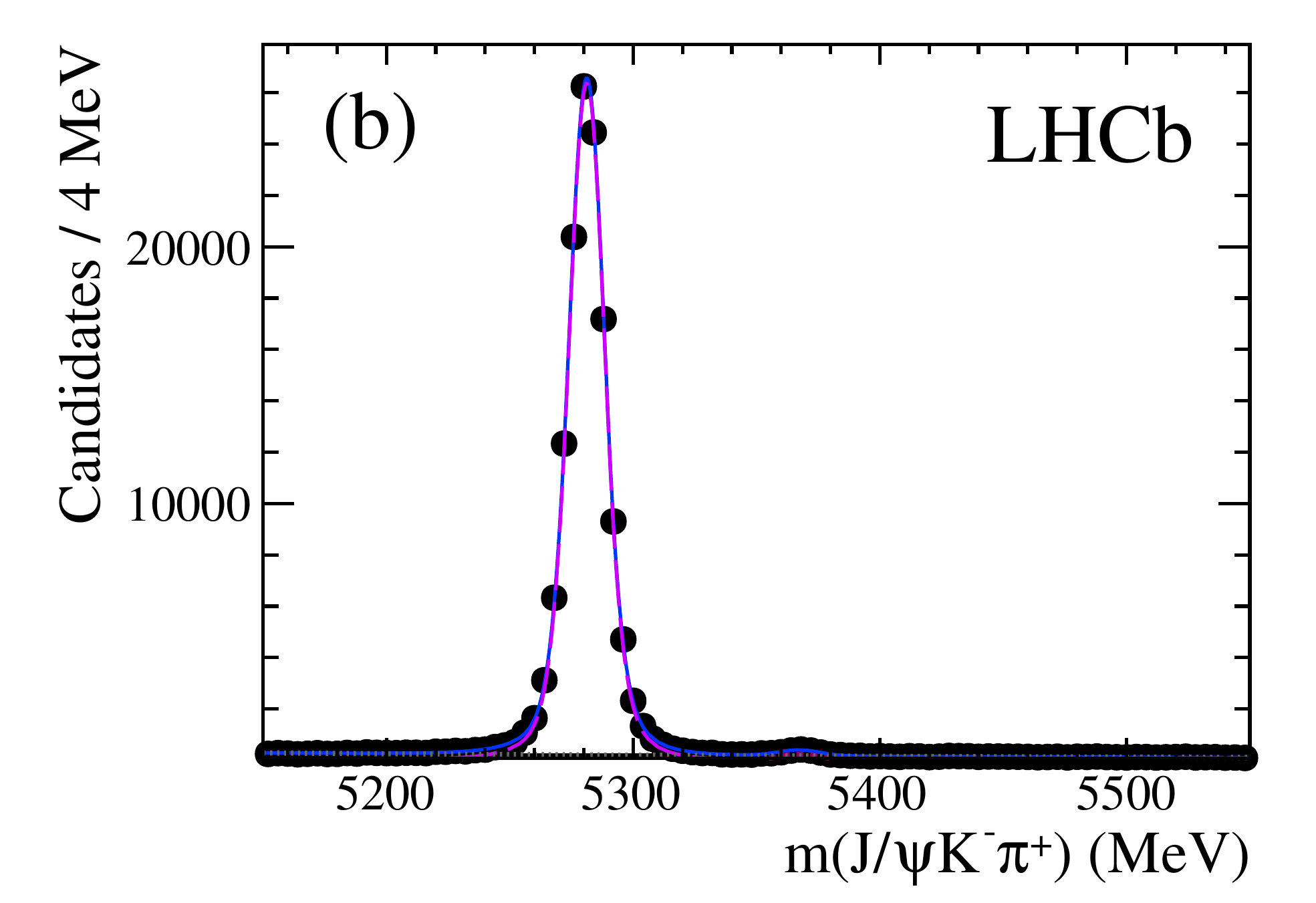}
        \caption{\small Invariant mass distributions of selected (a) $\jpsi\pipi$ and (b) $\jpsi\Km\pim$ candidates. The solid (blue) curves show the total fits, the long dashed (purple) curves show the respective $\Bsb\to\jpsi f_0(980)$ and $\Bdb \to \jpsi \Kstarzb(892)$ signals, and the dotted (grey) curve shows the combinatorial background. In (a) the short dashed (light blue) curve shows the \decay{\Bdb}{\jpsi\pipi} background and the dash dotted (green) curve shows the \decay{\Bdb}{\jpsi}{\Km\pip} reflection. In (b) the short dashed (pink) curve shows the \decay{\Bsb}{\jpsi\Km\pip} background.}
        \label{fig:massf0}
\end{center}
\end{figure}

The analysis uses the same selection criteria used to measure $\phi_s$ in $\Bsb\to\jpsi\pip\pim$ decays~\cite{LHCb:2012ad}. Events are triggered by the $\jpsi\to\mu^+\mu^-$ decay and a Boosted Decision Tree (BDT) is used to set the $\Bsb\to\jpsi\pip\pim$  selection requirements. The same trigger and BDT is used to select 
$\Bdb\to\jpsi\Kstarzb(892)$ events, except for the hadron identification that is applied independently of the BDT. Further selections of $\pm90$ MeV around the  $f_0(980)$ mass and $\pm100$ MeV around $\Kstarzb(892)$ mass~\cite{PDG}  are applied. The time-integrated fits to the $\jpsi f_0(980)$ and $\jpsi\Kstarz(892)$ invariant mass spectra are shown in Fig.~\ref{fig:massf0}. 

\begin{figure}
\begin{center}
            \includegraphics[width=0.60\textwidth]{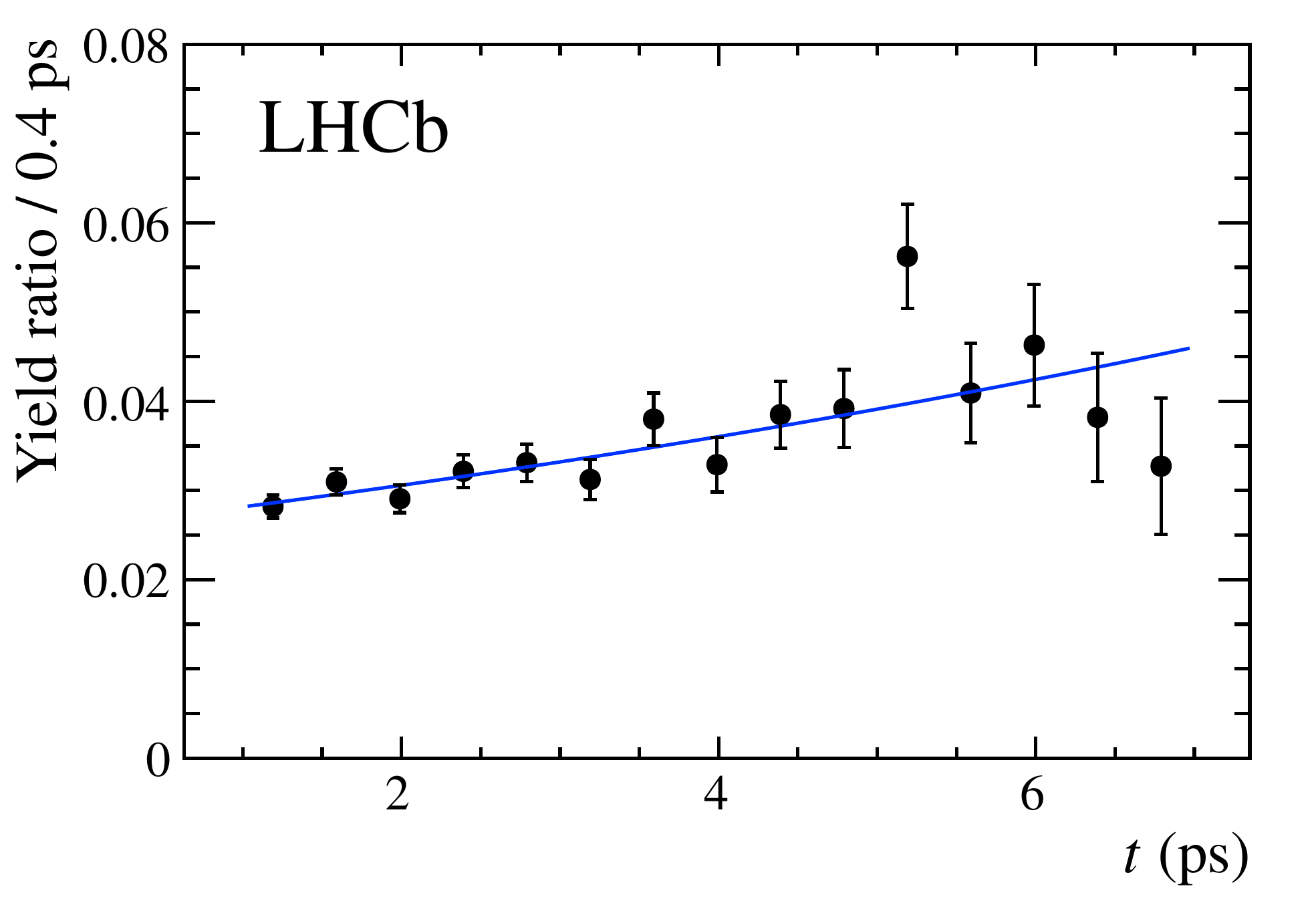}
        \caption{\small Decay time ratio between $\Bsb\to\jpsi f_0(980)$ and $\Bdb \to \jpsi \Kstarzb(892)$, and the fit for~$\Delta_{\jpsi f_0(980)}$.}
        \label{fig:f0ratio}
\end{center}
\end{figure}

The effective lifetime is measured from the variation of the ratio of the $B$ meson yields in bins of decay time: $R(t)= R(0)e^{-t(1/\tau_{\jpsi f_0(980)} - 1/\tau_{\jpsi\Kstarzb})} = R(0)e^{-t\Delta_{\jpsi f_0(980)}}$, where the width difference $\Delta_{\jpsi f_0(980)} = 1/\tau_{\jpsi f_0(980)} - 1/\tau_{\jpsi\Kstarzb}$. The decay time ratio between $\Bsb\to\jpsi f_0(980)$ and $\Bdb\to\jpsi\Kstarzb(892)$ with the fitted probability density function (PDF) $\Delta_{\jpsi f_0(980)}$ superimposed is shown in Fig.~\ref{fig:f0ratio}.  Making use of a binned fit, the reciprocal lifetime difference is determined to be $\Delta_{\jpsi f_0(980)}=-0.070\pm 0.014~\rm ps^{-1}$, where the uncertainty is statistical only. Taking $\tau_{\jpsi \Kstarzb}$ to be the mean $\Bdb$ lifetime $1.519\pm 0.007$ ps~\cite{PDG}, it is possible to determine $\tau_{\jpsi f_0(980)}= 1.700 \pm 0.040  \pm0.026$ ps~\cite{Aaij:2012nta}. The main systematic contribution is due to non perfect cancellation between the acceptance corrections in the two channels. Interpreting this as the lifetime of the heavy $\Bsb$ meson eigenstate, with an additional source of systematic uncertainty due to possible non-zero value of $\phi_s$, we obtain $\Gamma_H=0.588\pm0.014\pm0.009~\rm ps^{-1}$~\cite{Aaij:2012nta}.

\section{Precision measurement of the $\Lb$ baryon lifetime }
\noindent
The ratio of lifetimes of $\Lb$ baryon and $\Bdb$ meson, $\tau_{\Lb}/\tau_{\Bzb}$, is determined  using a data sample corresponding to 1.0~fb$^{-1}$ of integrated luminosity accumulated by the LHCb experiment in 7~TeV center-of-mass energy $pp$ collisions. The \Lb baryon is detected in the $\jpsi p K^-$ decay mode, while the \Bzb meson is found in $\jpsi \pi^+K^-$ decays. This \Lb decay mode has not been observed before.\footnote{Measurement of the branching fraction is under study.} On the other hand, the \Bzb decay is well known,  and we impose the further requirement that the invariant mass of the $\pi^+K^-$ combination be within $\pm$100~MeV of the $\Kstarzb(892)$ mass, in order to simplify the simulation and reduce systematic uncertainties. These decays have  the same decay topology into four charged tracks, thus facilitating the cancellation of uncertainties. The method is similar to that used in the  effective lifetime measurement of $\Bsb$ meson in $\jpsi f0(980)$ final state~\cite{Aaij:2012nta}, described above.

\begin{figure}[t]
\begin{center}
    \includegraphics[width=0.60\textwidth]{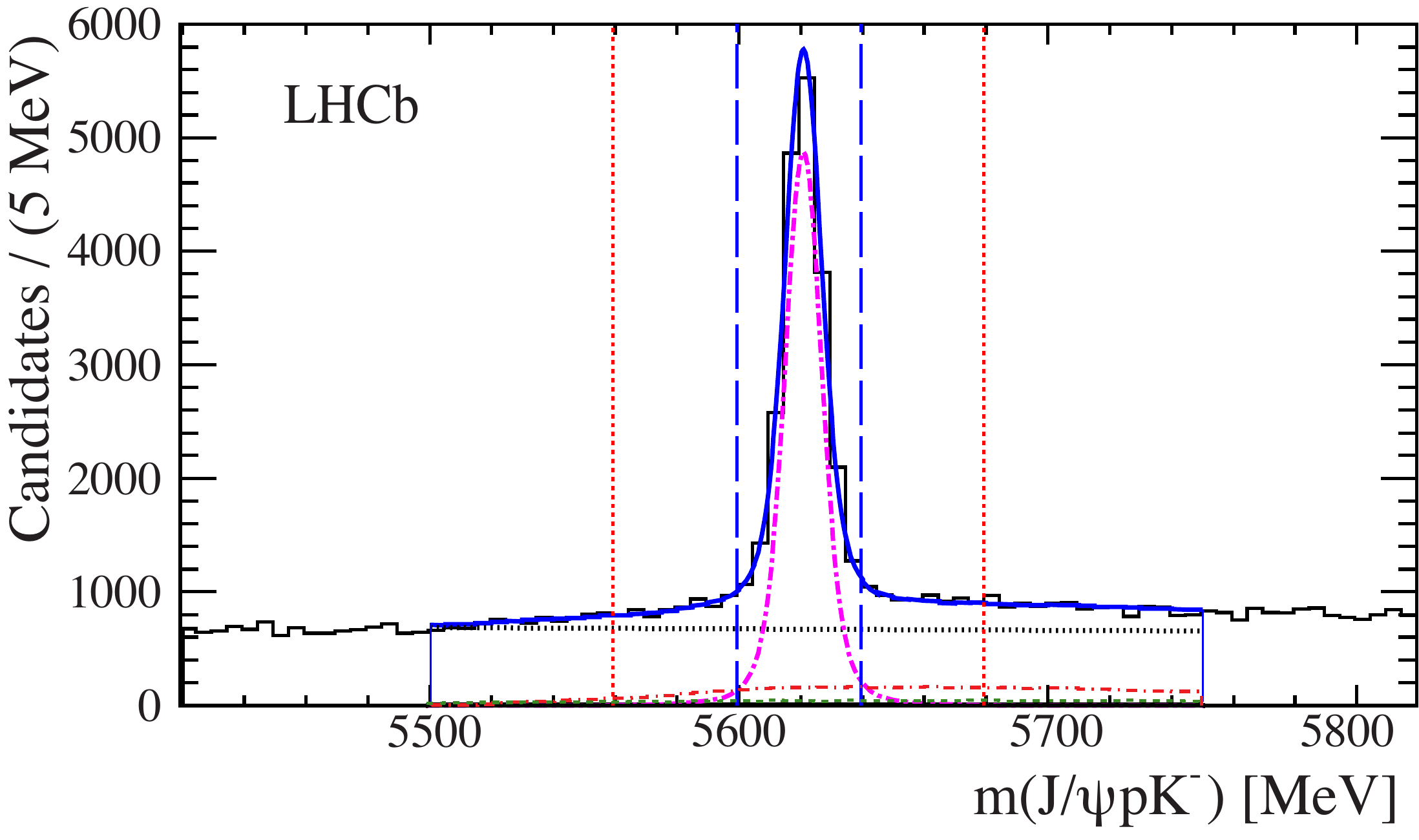}%
\end{center}\label{fig:psi-pK-mass}
\vskip -0.7cm
\caption{\small  Invariant mass spectrum of $\jpsi pK^-$ combinations. The signal region is between the vertical long dashed (blue) lines.  The sideband regions extend from the dotted (red) lines to the edges of the plot. The fit to the data between 5500 and 5750 MeV  is  also shown by the (blue) solid curve, with the $\Lb$ signal shown by the dashed-dot (magenta) curve. The  (black) dotted line is the combinatorial background and  $\Bsb\to \jpsi \KpKm$ and $\Bdb \to \jpsi \pi^+K^-$ reflections are shown with the (red) dashed-dot-dot  and (green) dashed shapes, respectively.}
\end{figure}

\begin{figure}[h!bt]
\begin{center}
\includegraphics[width=0.60\textwidth]{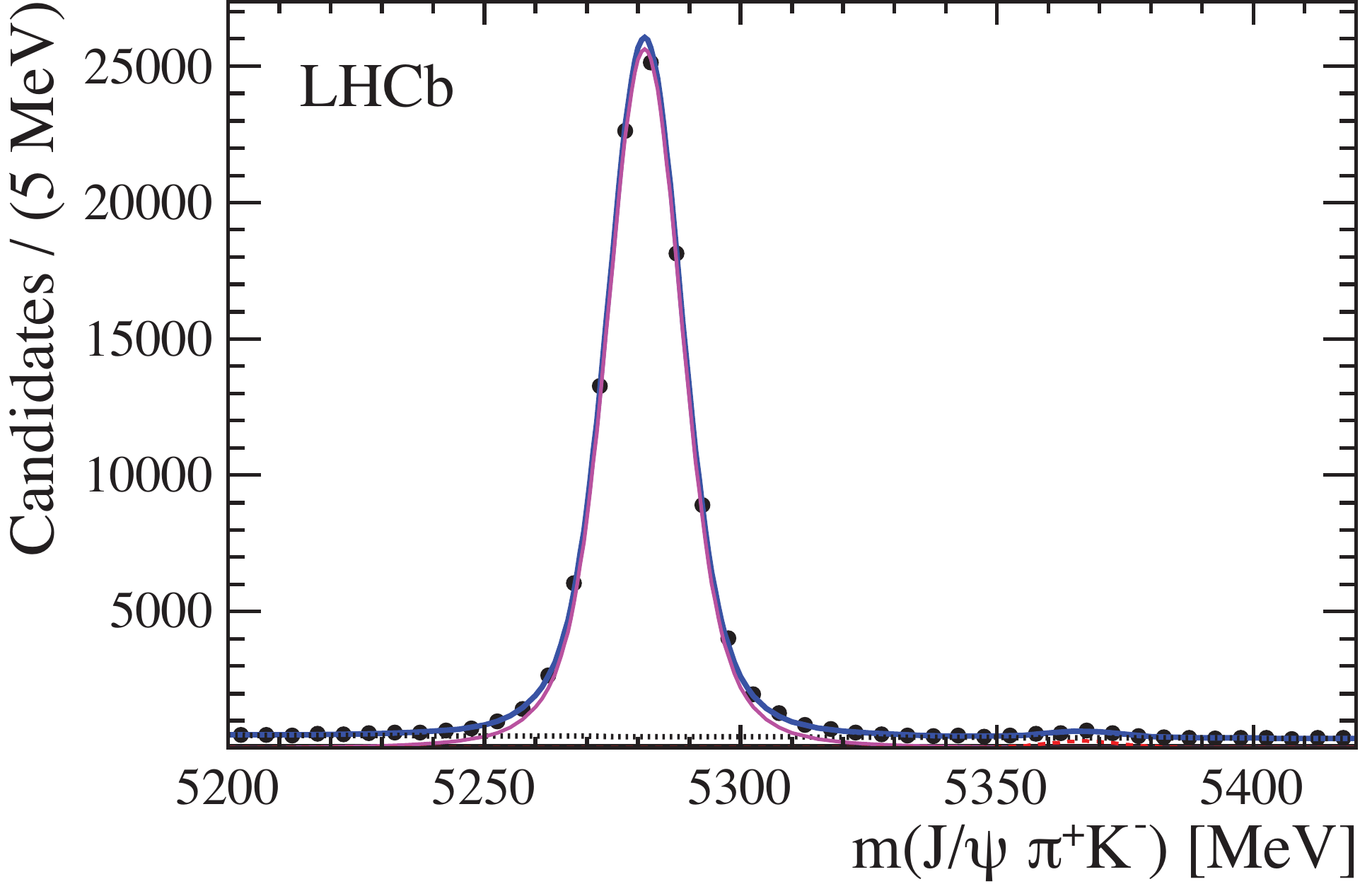}
\end{center}\label{fig:Bd2JpsiKst}
\vskip -0.4cm
\caption{\small Fit to the invariant mass spectrum of $\jpsi \pi^+K^-$ combinations with $\pi^+K^-$ invariant mass within $\pm$100~MeV of the $\Kstarzb$ mass. The $\Bdb$ signal is shown by the (magenta) solid  curve,  the combinatorial background by the (black) dotted line, the  $\Bs\to \jpsi \pi^+K^-$ signal  by the (red) dashed  curve,  and the total by the (blue) solid  curve.}
\end{figure}

\begin{figure}[htb]
\begin{center}
    \includegraphics[width=0.60\textwidth]{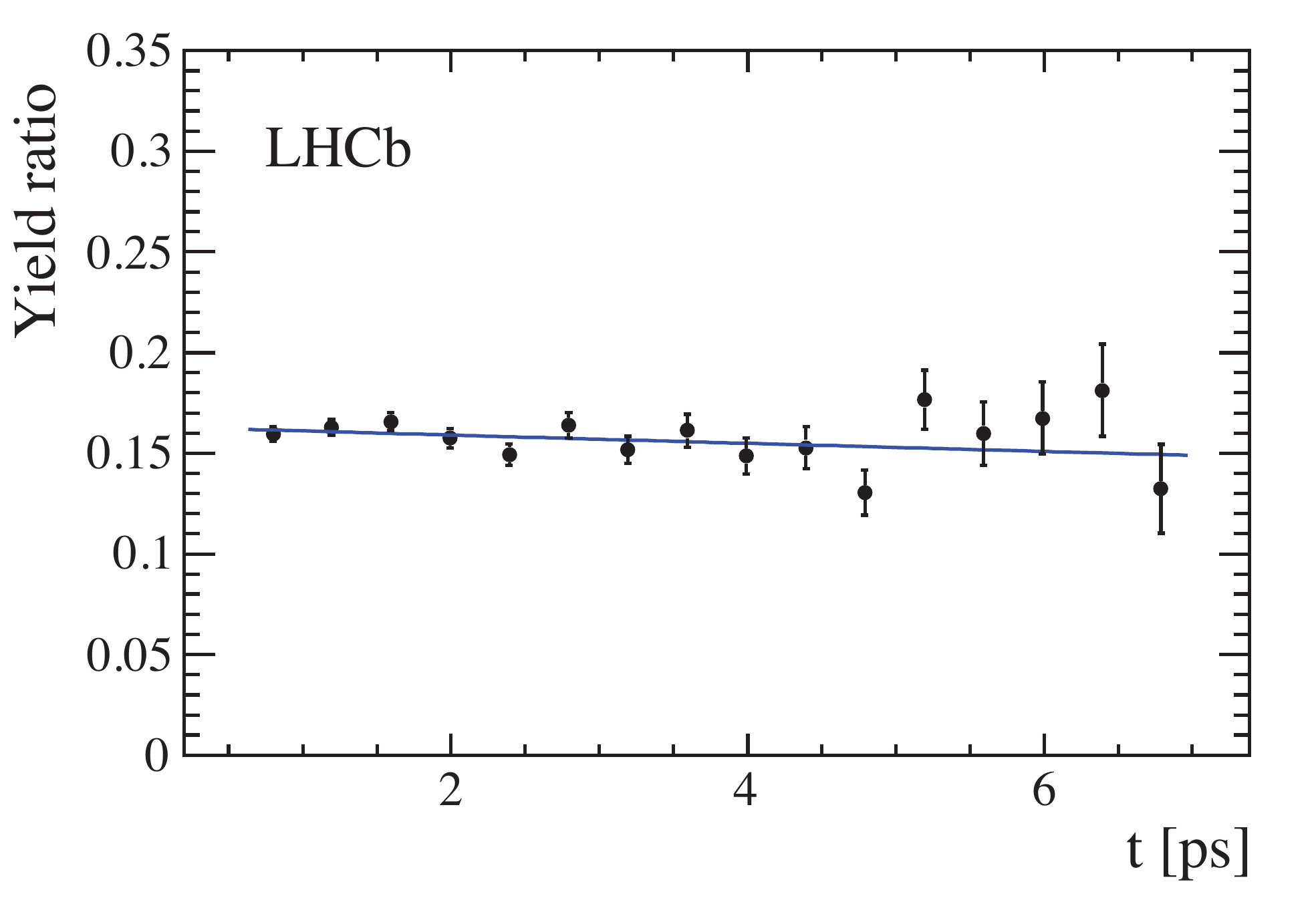}%
\end{center}\label{fig:yield_ratio}
\vskip -0.5cm
\caption{\small Yield ratio of $\Lb \to \jpsi pK^-$ to $\Bdb \to \jpsi \Kstarzb(892)$ events fitted as a function of decay time.}
\end{figure}

The  mass distributions of $J/\psi p K^-$ and $\jpsi \pi^+K^-$ combinations are  shown in Fig.~\ref{fig:psi-pK-mass} and Fig.~\ref{fig:Bd2JpsiKst}, respectively.
The decay time ratio distribution fitted with the PDF of $\Delta_{\jpsi p\Km}$ is shown in Fig.~\ref{fig:yield_ratio}.
The fitted value of the reciprocal lifetime difference is $\Delta_{\jpsi p\Km} = 16.4 \pm 8.2 \pm 4.4~ \rm ns^{-1}$ and the ratio of lifetimes, $\tau_{\Lb}/\tau_{\Bzb}$, is determined to be $\tau_{\Lb}/\tau_{\Bzb}=0.976\pm0.012\pm0.006$~\cite{Aaij:2013oha}. The ratio is equal to within a few percent, as the original advocates of the HQE claimed~\cite{Uraltsev:1998bk, Neubert:1996we},  without any need to find additional corrections.  Adding both uncertainties in quadrature, the lifetimes are consistent with being equal at the level of 1.9 standard deviations; thus we do not exclude that the \Lb baryon has a longer lifetime than the \Bzb meson. Using the world average measured value for the \Bzb lifetime~\cite{PDG} we determine
$\tau_{\Lb}= 1.482 \pm 0.018 \pm 0.012$~ ps~\cite{Aaij:2013oha}. The main systematic uncertainty is due to the change of the default fit range $0.6-7.0$ ps to $1.0-7.0$ ps. 

\section{Summary}
\noindent
Using $1.0~\rm fb^{-1}$ of data 
collected by the LHCb detector we present world best measurements of $\Bs$ mixing parameters, the width difference, $\Delta \Gamma_s$, the average width, $\Gamma_s$, $\Bs$ effective lifetime and the precision measurement of $\Lb$ baryon lifetime. They are in good agreement with the corresponding experimental world averages and HQE predictions. It is to be noted that all the measurements presented in this proceeding are from about 1/3 the of
the current LHCb dataset. We expect significant improvements to these results in the near future as more data are analyzed.

\section*{Acknowledgements}
\noindent
The author would like to thank  Sheldon Stone, Greig Cowan and Diego Martinez Santos for diligently reviewing and providing valuable feedback to the final manuscript. This work is supported by the U.S. National Science Foundation.

\end{document}